\documentclass[10pt]{iopart}
\usepackage{psfig}
\begin{document}
\eqnobysec
%
%%%%%%%%%%%%%%%%%%%%%%%%%%%%%%%%%%%%%%%%%%%%%%%%%%%%%%%%%%%%%%%%%%%%%
%%%%%%%%%%%%%%%%%%%%%%%%%%%%%%%%%%%%%%%%%%%%%%%%%%%%%%%%%%%%%%%%%%%%%

\title[Mean first-passage and residence times]
{Mean first-passage and residence times of 
random walks on asymmetric disordered chains}

\author{Pedro A Pury\dag\
\footnote[4]{pury@famaf.unc.edu.ar}
and Manuel O C\'aceres\ddag\ \S 
}

\address{\dag\
Facultad de Matem\'atica, Astronom\'\i a y F\'\i sica,
Universidad Nacional de C\'ordoba,
Ciudad Universitaria, 5000 C\'ordoba, Argentina}

\address{\ddag\
Centro At\'omico Bariloche and Instituto Balseiro,
%Comisi\'on Nacional de Energ\'\i a At\'omica and
%Universidad Nacional de Cuyo,
8400 San Carlos de Bariloche, R\'\i o Negro, Argentina}

\address{\S\
Consejo Nacional de Investigaciones Cient\'\i ficas
y T\'ecnicas, Argentina}

%%%%%%%%%%%%%%%%%%%%%%%%%%%%%%%%%%%%%%%%%%%%%%%%%%%%%%%%%%%%%%%%%%%%

\begin{abstract}
An algebraic derivation is presented which yields the exact
solution of the mean first-passage and mean residence times
of a one-dimensional asymmetric random walk for quenched disorder.
Two models of disorder are analytically treated.
Both, absorbing-absorbing and reflecting-absorbing boundaries 
are considered.
Particularly, the interplay between asymmetry and disorder
is studied.
\end{abstract}

%%%%%%%%%%%%%%%%%%%%%%%%%%%%%%%%%%%%%%%%%%%%%%%%%%%%%%%%%%%%%%%%%%%%%
%%%%%%%%%%%%%%%%%%%%%%%%%%%%%%%%%%%%%%%%%%%%%%%%%%%%%%%%%%%%%%%%%%%%%

\submitto{Journal of Physics A: Received 21 October 2002,
in final form 9 January 2003}

%\maketitle

%%%%%%%%%%%%%%%%%%%%%%%%%%%%%%%%%%%%%%%%%%%%%%%%%%%%%%%%%%%%%%%%%%%%%
\section{Introduction}
\label{sec:Intro}
%%%%%%%%%%%%%%%%%%%%%%%%%%%%%%%%%%%%%%%%%%%%%%%%%%%%%%%%%%%%%%%%%%%%%

The problems of the first-passage time (FPT)~\cite{Red01,Kam92} 
and the residence time (RT)~\cite{BZA98} are very important issues 
in random walk theory. Moreover, several properties of diffusion
and transport in disordered systems are based in this concepts.
Here, we do not want to present a survey of the enormous literature
in the field of mean first-passage time (MFPT) and mean residence
time (MRT) of random walks, nevertheless, we wish to single out the
works involved with analytical or exact results, mainly in
one-dimensional disordered systems.
Goldhirsch and Gefen~\cite{GG86} developed an analytical method 
for calculating MFPT for branched networks such as finite segments 
with dangling bonds and loops. The method is based on the generating
function and was generalized for biased walks~\cite{GG87}.
Extensions of the generating function method were done for
analyzing the probability distribution function of FPT~\cite{NG87}
and the current autocorrelation function~\cite{GN87}.
Later on, the generating function method was used for random
one-dimensional chains~\cite{NG88,NG90}, particularly for
the Sinai problem~\cite{Sinai}.
Explicit expressions for the MFPT, in terms of the basic
jump probabilities for discrete time random walk with a reflecting
boundary were obtained independently by Van den Broeck~\cite{VdB89},
Le Doussal~\cite{LeD89}, and Murthy and Kehr~\cite{MK89} by different
methods. It is interesting to remark that Gardiner~\cite{Gar85}
had previously reported explicit MFPT formulae.
An exact solution of the generating function for the first-passage
probability was presented by Raykin~\cite{Ray93} using enumerative
conbinatorics for summing up over trajectories of the random 
walker~\cite{CCM02}.
The distribution of escape probabilities was computed exactly 
by Sire~\cite{Sir99} and the exact renormalization group analysis 
was performed by Le Dousal \etal~\cite{LMF99}.
In the last few years, one of the main applications of the exact
expressions for MFPT and MRT in one-dimensional lattices was
the study of exciton migration in treelike dendrimers
(light harvesting antennae)~\cite{BK98,RSM02}.

A successful perturbative theory for survival statistics in 
disordered chains is the finite effective medium approximation 
(FEMA)~\cite{HC90}.
Extensions of FEMA to biased media~\cite{Pury94} and periodically
forced boundary conditions~\cite{CMO+97} were presented.
A unified framework for the FPT and RT statistics in finite
disordered chains with bias was also presented by the authors in
reference~\cite{Pury02}, where exact equations for the
quantities averaged over disorder were obtained for both problems
and its solutions up to first order in the bias parameter were
constructed retaining the full dependence on the system's size
and the initial condition.

In this paper, we obtain explicit analytical expressions for
the MFPT and MRT of random walks on a one-dimensional lattice
for a quenched realization of disorder.
Then, we consider two models for the disorder in the hopping
transitions and average the expressions, in each case, on the
realizations of disorder.  
The outline of the paper is as follow. The starting point of our
formulation is given in section~\ref{sec:prob}.
In sections~\ref{sec:MFPT} and~\ref{sec:MRT} we discuss the algebraic
method that allows us obtain the exact dependence of the MFPT and
the MRT on the set of transition probabilities.  The biased
nondisordered chain is treated in section~\ref{sec:order}, whereas
two models of disorder with asymmetric transition probabilities are
analyzed in sections~\ref{sec:weak} and~\ref{sec:mult}. The effects
on MFPT and MRT of cutting the chain, at a reflecting site, are
considered in section~\ref{sec:cut}.
Finally, in section~\ref{sec:fin}, we briefly summarize the principal
results of our study.

%%%%%%%%%%%%%%%%%%%%%%%%%%%%%%%%%%%%%%%%%%%%%%%%%%%%%%%%%%%%%%%%%%%%%
\section{Survival and residence probabilities}
\label{sec:prob}
%%%%%%%%%%%%%%%%%%%%%%%%%%%%%%%%%%%%%%%%%%%%%%%%%%%%%%%%%%%%%%%%%%%%%

We start considering a random walk on a discrete one-dimensional
lattice with nearest neighbor hopping; jumping from site $n$
to site $n+1$ with transition probability $w_n^+$,
or to site $n-1$ with transition probability $w_n^-$.
In this manner, the walker has a sojourn probability
$1 - w_n^+ - w_n^-$ per unit time at site $n$.
The conditional probability, $P(m,t|n)$, of finding the walker 
at site $m$ at time $t$, given that it was initially at site $n$,
satisfies a Markovian master equation for a given realization
of the set $\{w_j^{\pm}\}$.

We are concerned with the survival and residence probabilities
in the finite interval $D = [-M, L]$ on the chain. 
The first is the probability, $S_n(t)$, of remaining in $D$ 
at time $t$, without exiting, if the walker initially began 
at site $n \in D$. The second is the probability, $R_n(t)$, 
of finding the particle within the domain $D$ at time $t$, 
given that it initially was at site $n$ (not necessary in $D$).
The dynamical evolution of both quantities follows from the backward
master equation~\cite{Pury02}
\begin{equation}
\partial_t F_{n}(t) =
w^+_{n} \,\left( F_{n+1}(t) - F_{n}(t) \right) +
w^-_{n} \,\left( F_{n-1}(t) - F_{n}(t) \right)
\;,
\label{EqFt}
\end{equation}
where
\begin{equation}
F_{n}(t) = \sum_{m \in D} \,P(m,t|n) \;.
\label{Ft}
\end{equation}
The survival probability is the solution of the
equation~(\ref{EqFt}) with the initial condition $S_{n}(t=0)=1$,
for all $n \in D$, considering the boundary conditions
$S_{-(M+1)}(t) = S_{{L+1}}(t) = 0$, for all $t$.
On the other hand, the residence probability is the solution
of the equation~(\ref{EqFt}) with the initial condition
$R_{n}(t=0) = 1$ if $n \in D$, or $0$ otherwise,
fulfilling the boundary condition $R_{n}(t) \rightarrow 0$, 
for $|n| \rightarrow \infty$ for all finite $t$.

Finally, MFPT and MRT can be obtained from the asymptotic limit
of the Laplace transformed (denoted by hats) survival and residence
probabilities, respectively~\cite{Pury02},
\numparts
\begin{eqnarray}
T_{n} = \lim_{s \rightarrow 0} \hat{S}_{n}(s) \;,
\label{T_Ss}
\\
\tau_{n} = \lim_{s \rightarrow 0} \hat{R}_{n}(s) \;.
\label{tau_Rs}
\end{eqnarray}
\endnumparts
%

%%%%%%%%%%%%%%%%%%%%%%%%%%%%%%%%%%%%%%%%%%%%%%%%%%%%%%%%%%%%%%%%%%%%%
\subsection{Mean first-passage time}
\label{sec:MFPT}
%%%%%%%%%%%%%%%%%%%%%%%%%%%%%%%%%%%%%%%%%%%%%%%%%%%%%%%%%%%%%%%%%%%%%

\begin{figure}
\begin{center}
\psfig{file=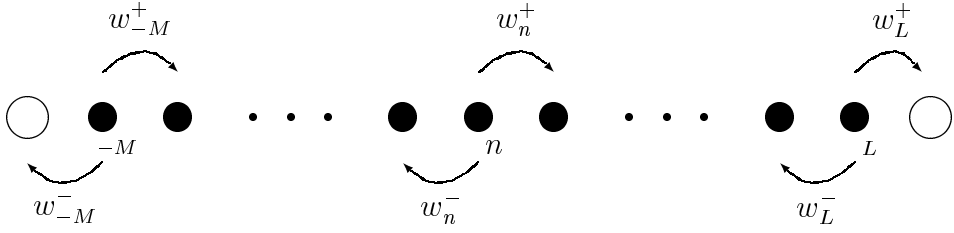}
\end{center}
\caption{
Graphical illustration of the definition of the probability
transitions. The boundary conditions for the first-passage
time problem are perfect tramps (\opencircle) in the segment
extremes.
}
\label{FPT-Fig}
\end{figure}

In the first-passage time problem the random walker only jumps
inside a finite interval with absorbing ends, as shown in
figure~\ref{FPT-Fig}. The boundary can be simply modeled by setting
$w^+_{-(M+1)} = w^-_{L+1} = 0$ (i.e., the walker cannot jump back
into the interval once it has been tramped on $-(M+1)$ or $L+1$).
From the Laplace transform of the evolution
equation~(\ref{EqFt}), using that $S_{n}(0)=1$
for all $n \in D$, and taking the limit of 
equation~(\ref{T_Ss}), we obtain the corresponding 
equation for the MFPT
\begin{equation}
w^+_{n} \,\left( T_{n+1} - T_{n} \right) +
w^-_{n} \,\left( T_{n-1} - T_{n} \right) = -1
\,, \qquad
\forall \;n \in D.
\label{eq_T}
\end{equation}
This equation must be suplemented with the conditions:
$T_{-(M+1)} = T_{L+1} = 0$.
Equation~(\ref{eq_T}) is a three-term recursion formula.
To get a two-term recursion relation, which is simpler
to analyze, following Gardiner~\cite{Gar85} we make 
the substitution
\begin{equation}
\Delta_n = T_{n+1} - T_n \;.
\label{Delta}
\end{equation}
This yields the equation
$w^+_{n} \,\Delta_n - w^-_{n} \,\Delta_{n-1} = -1$,
and results in $\Delta_L = - T_L$ and $\Delta_{-(M+1)} = T_{-M}$.
Assuming that $w^+_{n} \neq 0$ for all $n \in D$,
from these conditions we immediately obtain
\begin{equation}
\Delta_n = \frac{w^-_n}{w^+_n} \,\Delta_{n-1}
- \frac{1}{w^+_n} \;,
\label{auxD1}
\end{equation}
\begin{equation}
\sum_{n \in D} \Delta_n = -T_{-M} \;.
\label{auxD2}
\end{equation}
Starting from site $-M$ and applying recursively~(\ref{auxD1}),
we obtain
\begin{equation}
\Delta_n = \prod_{j=-M}^{n} \;\frac{w^-_j}{w^+_j} \,T_{-M}
- \frac{1}{w^+_n}
- \frac{1}{w^+_n} \,
\sum_{i=-M}^{n-1} \prod_{j=i}^{n-1}
\;\frac{w^-_{j+1}}{w^+_j}
\;.
\label{Delta-n}
\end{equation}
Note that the last term only runs for $n \geq -M+1$.
Using~(\ref{Delta-n}) in~(\ref{auxD2}), immediately results in
\begin{equation}
\fl
T_{-M}
\left(
1 + \sum_{k=-M}^{L} \prod_{j=-M}^{k}
\;\frac{w^-_{j}}{w^+_j}
\right)
=
\sum_{k=-M}^{L} \,\frac{1}{w^+_k} +
\sum_{k=-M+1}^{L} \,\frac{1}{w^+_k} \,
\sum_{i=-M}^{k-1} \prod_{j=i}^{k-1}
\;\frac{w^-_{j+1}}{w^+_j}
\;.
\label{T-M}
\end{equation}
Finally, writing
$T_n = T_{-M} + \displaystyle\sum_{k=-M}^{n-1} \,\Delta_k$
and substituting in according to equations~(\ref{Delta-n})
and~(\ref{T-M}), we get
\begin{eqnarray}
\fl
T_n =
\frac
{
\displaystyle
{1+\sum_{k=-M}^{n-1}\prod_{j=-M}^{k}\;\frac{w^-_j}{w^+_j}}}
{
\displaystyle
{1 +\sum_{k=-M}^{L} \prod_{j=-M}^{k}\;\frac{w^-_j}{w^+_j}}}
\left(
\displaystyle\sum_{k=-M}^{L} \,\frac{1}{w^+_k} +
\displaystyle\sum_{k=-M+1}^{L} \,\frac{1}{w^+_k} \,
\sum_{i=-M}^{k-1} \prod_{j=i}^{k-1}
\;\frac{w^-_{j+1}}{w^+_j}
\right)
\nonumber \\
\lo-
\left(
\displaystyle\sum_{k=-M}^{n-1} \,\frac{1}{w^+_k} +
\displaystyle\sum_{k=-M+1}^{n-1} \,\frac{1}{w^+_k} \,
\sum_{i=-M}^{k-1} \prod_{j=i}^{k-1}
\;\frac{w^-_{j+1}}{w^+_j}
\right)
\label{MFPT_bond}
\;.
\end{eqnarray}
This expression can be additionally recast as
\begin{eqnarray}
\fl
T_n =
\frac
{\displaystyle
{1+\sum_{k=-M}^{n-1}\prod_{j=-M}^{k}\;\frac{w^-_j}{w^+_j}}}
{\displaystyle
{1 +\sum_{k=-M}^{L} \prod_{j=-M}^{k}\;\frac{w^-_j}{w^+_j}}}
\left(
\displaystyle\sum_{k=-M}^{L} \,\frac{1}{w^+_k} +
\displaystyle\sum_{k=-M}^{L-1} \,\frac{1}{w^+_k} \,
\sum_{i=k+1}^{L} \prod_{j=k+1}^{i} \;\frac{w^-_j}{w^+_j}
\right)
\nonumber \\
\lo-
\left(
\displaystyle\sum_{k=-M}^{n-1} \,\frac{1}{w^+_k} +
\displaystyle\sum_{k=-M}^{n-2} \,\frac{1}{w^+_k} \,
\sum_{i=k+1}^{n-1} \prod_{j=k+1}^{i} \;\frac{w^-_j}{w^+_j}
\right)
\;,
\label{MFPT:aa}
\end{eqnarray}
for $n \in D$, and where the sums whose upper limit is
$n - \alpha$ run only for $n \geq -M + \alpha$.
The above equation for the MFPT is an exact expression for
quenched disorder, given that it contains explicitly the
full dependence on the basic jump transitions $\{w_j^{\pm}\}$.

Let us substitute, in (\ref{MFPT:aa}), $w^+_k = w^-_k = w_k$
for all $k$. This corresponds to a symmetrical random walk and
we obtain
\begin{eqnarray}
%\fl
T_n =& \frac{n+M+1}{L+M+2}
\left( \sum_{k=-M}^{L} \frac{1}{w_k}
+ \sum_{k=-M}^{L-1} \frac{L-k}{w_k} \right)
\nonumber\\
%\lo
&-
\left( \sum_{k=-M}^{n-1} \frac{1}{w_k}
+ \sum_{k=-M}^{n-2} \frac{n-k-1}{w_k} \right)
\;.
\label{symm:aa}
\end{eqnarray}
%

%%%%%%%%%%%%%%%%%%%%%%%%%%%%%%%%%%%%%%%%%%%%%%%%%%%%%%%%%%%%%%%%%%%%%
\subsection{Mean residence time}
\label{sec:MRT}
%%%%%%%%%%%%%%%%%%%%%%%%%%%%%%%%%%%%%%%%%%%%%%%%%%%%%%%%%%%%%%%%%%%%%

\begin{figure}
\begin{center}
\psfig{file=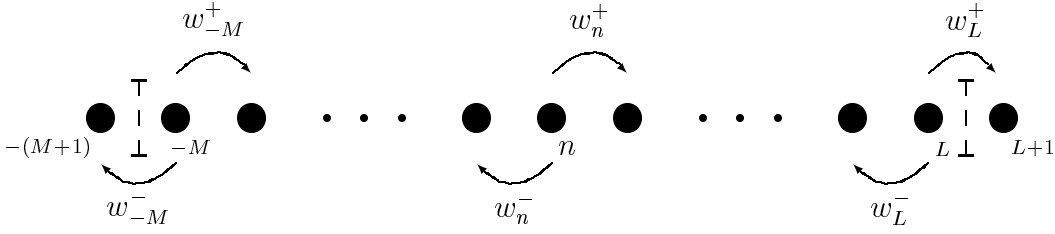}
\end{center}
\caption{
Schematic representation of the residence time problem.
The region of interest is the segment between the vertical
dashed lines.
}
\label{RT-Fig}
\end{figure}

In the residence time problem, the walker jumps on the unbounded
chain, but we compute the probability of finding the walker in the 
finite region $D$, as shown in figure~\ref{RT-Fig}.
Particularly, we are concerned with the mean time that 
the walker spends in $D$.
From the Laplace transform of the evolution equation~(\ref{EqFt})
and taking the limit of equation~(\ref{tau_Rs}),
we obtain the corresponding equation for the MRT
\begin{equation}
w^+_{n} \,\left( \tau_{n+1} - \tau_{n} \right) +
w^-_{n} \,\left( \tau_{n-1} - \tau_{n} \right) = -R_n(t=0)
\,,
\label{eq_tau}
\end{equation}
where $R_{n}(t=0) = 1$ if $n \in D$, or $0$ otherwise.
For this problem, we assume the presence of a global bias that 
points to the right, i.e., $w^-_j/w^+_j < 1$ for all site $j$.
Thus, equation (\ref{eq_tau}) must be supplemented with the 
condition: $\lim_{n \rightarrow +\infty} \tau_n = 0$.
Equation~(\ref{eq_tau}) is also a three-term recursion formula.
Therefore, we make again the substitution
\begin{equation}
\Gamma_n = \tau_{n+1} - \tau_n \;.
\label{Gamma}
\end{equation}
This yields the equations
\numparts
\begin{eqnarray}
\Gamma_{n-1} = \frac{w^+_n}{w^-_n} \,\Gamma_{n}
\qquad \mbox{for } n < -M
\label{auxG2a}
\;,
\\
\Gamma_n = \frac{w^-_n}{w^+_n} \,\Gamma_{n-1}
\qquad \mbox{for } n > L
\label{auxG2b}
\;,
\end{eqnarray}
\begin{equation}
\Gamma_n = \frac{w^-_n}{w^+_n} \,\Gamma_{n-1}
- \frac{1}{w^+_n}
\qquad \mbox{for } -M \leq n \leq L \;.
\label{auxG1}
\end{equation}
\endnumparts
We are considering residence in a finite region,
then we must take the boundary conditions:
$\lim_{n \rightarrow \pm\infty} \Gamma_n = 0$.

Applying recursively equation~(\ref{auxG2a}), we get
\begin{equation}
\Gamma_k = \prod_{j=k+1}^{-(M+1)} \frac{w^+_j}{w^-_j}
\, \Gamma_{-(M+1)} \qquad \mbox{for }k \leq -(M+2) \;.
\label{G<-(M+1)}
\end{equation}
Given that we have assumed $w^+_j/w^-_j > 1$, the condition
$\lim_{k \rightarrow -\infty} \Gamma_k = 0$, imposes that
$\Gamma_{-(M+1)} = 0$. Therefore, we immediately obtain
$\Gamma_k = 0$ for $k \leq -(M+1)$.
Thus, from equation~(\ref{Gamma}) we get
\begin{equation}
\tau_k = \tau_{-M} \qquad \mbox{for } k < -M.
\label{tau<-M}
\end{equation}
In a similar way, from equation~(\ref{auxG2b}), we get
\begin{equation}
\Gamma_k = \prod_{j=L+1}^{k} \frac{w^-_j}{w^+_j}
\, \Gamma_{L} \qquad \mbox{for }k \geq L+1 \;,
\label{G>L}
\end{equation}
and the assumption $w^-_j/w^+_j < 1$ guarantees that
$\lim_{k \rightarrow +\infty} \Gamma_k = 0$, for any
arbitrary $\Gamma_{L}$.
On the other hand, from equation~(\ref{auxG1}), using that 
$\Gamma_{-(M+1)} = 0$, we obtain
\numparts
\begin{eqnarray}
\Gamma_{-M} & = -\frac{1}{w^+_{-M}} \;,
\label{G_-M}
\\
\Gamma_k &= -\frac{1}{w^+_{k}}
\left(
1 + \sum_{i=-M}^{k-1} \prod_{j=i}^{k-1} \frac{w^-_{j+1}}{w^+_j}
\right)
\nonumber \\
&= -\frac{1}{w^+_{k}} - \sum_{i=-M}^{k-1} \frac{1}{w^+_i}
\prod_{j=i+1}^{k} \frac{w^-_j}{w^+_j}
\qquad \mbox{for } -M < k \leq L \;.
\label{GinD}
\end{eqnarray}
\endnumparts

The boundary condition for the MRT 
when the bias points to the right,
$\lim_{n \rightarrow +\infty} \tau_n = 0$,
and equation~(\ref{Gamma}), allows us to write
$\tau_n = - \sum_{k=n}^{\infty} \Gamma_k$, 
and using equations~(\ref{G>L}) 
and~(\ref{GinD}) for $k=L$, we obtain
\begin{equation}
\fl
\tau_n = \left( \frac{1}{w_L^+}
+ \sum_{i=-M}^{L-1} \frac{1}{w_i^+}
\prod_{j=i+1}^{L} \frac{w_j^-}{w_j^+} \right) \,
\sum_{k=n}^{\infty} \prod_{j=L+1}^{k}
\frac{w_j^-}{w_j^+}
\qquad \mbox{for $n > L$},
\label{tau>L}
\end{equation}
and
\begin{equation}
\fl
\tau_L = \left( \frac{1}{w_L^+}
+ \sum_{i=-M}^{L-1} \frac{1}{w_i^+}
\prod_{j=i+1}^{L} \frac{w_j^-}{w_j^+} \right) \,
\left( 1 + \sum_{k=L+1}^{\infty} \prod_{j=L+1}^{k}
\frac{w_j^-}{w_j^+} \right)
\;.
\label{tau_L}
\end{equation}
From equation~(\ref{Gamma}), we can also write
$\tau_{-M} = \tau_L - \sum_{k=-M}^{L-1} \Gamma_k$,
and using equation~(\ref{GinD}), we get, after
introducing a change in the order of the sums,
\begin{equation}
\fl
\tau_{-M} = \tau_L + \sum_{k=-M}^{L-1} \frac{1}{w_k^+}
+ \sum_{k=-M}^{L-2} \frac{1}{w_k^+}
\sum_{i=k+1}^{L-1} \prod_{j=k+1}^{i} \frac{w_j^-}{w_j^+}
\;.
\label{tau_-M}
\end{equation}
Finally, writting
$\tau_n = \tau_{-M} + \sum_{k=-M}^{n-1} \Gamma_k$,
using equation~(\ref{GinD}), and doing the same change 
in the order of the sums, we obtain
\begin{equation}
\fl
\tau_{n} = \tau_{-M}
- \sum_{k=-M}^{n-1} \frac{1}{w_k^+}
- \sum_{k=-M}^{n-2} \frac{1}{w_k^+}
\sum_{i=k+1}^{n-1} \prod_{j=k+1}^{i} \frac{w_j^-}{w_j^+}
\qquad \mbox{for $-M < n < L$}.
\label{tauinD}
\end{equation}
Equations~(\ref{tau<-M}) and~(\ref{tau>L})--(\ref{tauinD}) are
the exact expressions for the MRT for quenched disorder, with
the full dependence on the set of transitions $\{w_j^{\pm}\}$.

For the unbiased random walk, the residence problem is not defined
as can be seen from equations~(\ref{tau>L})--(\ref{tau_L}), where 
the series diverges in the symmetrical case, i.e., $w^+_j = w^-_j$.

%%%%%%%%%%%%%%%%%%%%%%%%%%%%%%%%%%%%%%%%%%%%%%%%%%%%%%%%%%%%%%%%%%%%%
\section{Homogeneous chain}
\label{sec:order}
%%%%%%%%%%%%%%%%%%%%%%%%%%%%%%%%%%%%%%%%%%%%%%%%%%%%%%%%%%%%%%%%%%%%%

The homogeneous biased chain corresponds to the case
$w^+_j = a$, $w^-_j = b$. Introducing the bias parameter
$\gamma = b/a$, equation~(\ref{MFPT:aa}) can be easily written as
\begin{equation}
\fl
T_n = \displaystyle\frac{L+1-n}{a (1-\gamma)} -
      \frac{L+M+2}{a (1-\gamma)}
      \frac{\gamma^n-\gamma^{L+1}}{\gamma^{-(M+1)}-\gamma^{L+1}}
\qquad \mbox{for $-M \leq n \leq L$} .
\label{Tord:aa}
\end{equation}
The study of the drift and diffusive regimes of the MFPT follows
from equation~(\ref{Tord:aa}). Additional information about the
survival probability in the homogeneous chain was reported in 
references~\cite{Pury94} and~\cite{Pury02}.

For the MRT in the homogeneous chain, we get
\begin{equation}
\tau_n = \frac{1}{a} \left\{
\begin{array}{ll}
\displaystyle
\frac{L+M+1}{1-\gamma}
& \mbox{for $n < -M$} \\
\displaystyle
\frac{L-n}{1-\gamma} + \frac{1-\gamma^{n+M+1}}{(1-\gamma)^2}
& \mbox{for $-M \leq n \leq L$} \\
\displaystyle
\frac{1-\gamma^{L+M+1}}{(1-\gamma)^2} \;\gamma^{n-L}
& \mbox{for $n >  L$}  \;,
\end{array}
\right.
\label{tauord}
\end{equation}
where $0 < \gamma < 1$.
An expression for the residence probability in the homogeneous
chain and the limit regimes of equation~(\ref{tauord}) were given
in reference~\cite{Pury02}.

%%%%%%%%%%%%%%%%%%%%%%%%%%%%%%%%%%%%%%%%%%%%%%%%%%%%%%%%%%%%%%%%%%%%%
\section{Weak biased disordered chain}
\label{sec:weak}
%%%%%%%%%%%%%%%%%%%%%%%%%%%%%%%%%%%%%%%%%%%%%%%%%%%%%%%%%%%%%%%%%%%%%

Two prototype problems are defined in disordered one-dimensional
systems. The first one, the bond disorder or random-barrier problem,
corresponds to the situation in which the transfer rates associated
with the bond between two neighbor sites are symmetric,
$w_n^+ = w_{n+1}^-$.
In this kind of problem, the statistical properties of the system 
are given by the distribution of the random variable which describes
the transition rate in each bond.
The second class is the site disorder or random-trap problem.
Here, the hopping probabilities $w_n^{\pm}$ associated with each 
site are symmetric, $w_n^+ = w_{n+1}^-$~\cite{HC90}, and the
distribution of hopping rates in each site characterizes the 
model of disorder.
Physical realizations corresponding to each class of problems 
were summarized by Alexander \etal~\cite{ABSO81}.

In what follows, we consider site disordered chains. We assume 
that the hopping probabilities $w_n^{\pm}$ are strictly positive 
random variables, chosen independently from site to site and 
identically distributed. However, we are particularly interested 
in the effects of bias in the chain by external fields. Therefore,
we admit that the site transition probabilities are not necessarily 
symmetric in the sense that $w_n^+ \neq w_n^-$.
Thus, our first random-site model is defined by
$w^+_j = a + \xi_j$, $w^-_j = b + \xi_j$,
where $a$ and $b$ are positive constants, and $\{ \xi_j \}$ are
taken independent but identically distributed random variables with
$\langle \xi_j \rangle = 0$.
This form of jump transitions involves an ordered biased background
with an added random medium. The strength of the bias is given
by the ratio between $a$ and $b$ and the disorder is characterized
by the distribution of variables $\left\{ \xi_j \right\}$.
We introduce the parameter $\epsilon$ for bias strength by
$b/a = 1 - \epsilon$.
This selection of parameters allows us to focus our attention in
the small bias limit and to study the transition to the symmetric
diffusive behavior. 
For practical reasons, we can alternatively introduce this 
{\em additive} model by the constrain $w^-_j = w^+_j - a \epsilon$,
where the random variables $\{ w^+_j \}$ are taken independent and
identically distributed. Additionally, $w^+_j > a \epsilon$ 
for $\epsilon > 0$,  otherwise $w^+_j$ is strictly positive 
for $\epsilon < 0$. Thus, we can write
\begin{equation}
\frac{w^-_j}{w^+_j} = 1 - \frac{a \epsilon}{w^+_j} .
\label{w-/w+}
\end{equation}
We will find it useful to define the quantities
$\beta_k \equiv \left< \left( 1 / w^+_j \right)^k \right>$,
which we assume finite for all $k \geq 1$.
This class of disorder is known as
weak disorder~\cite{HC90,Pury94,CMO+97,Pury02}.
We also define a measure of the fluctuation of the disorder by
${\cal F} \equiv \left(\beta_2 - \beta_1^2 \right) / \beta_1^2$.
In this manner, we can write
\begin{equation}
\left< \frac{1}{w^+_j w^+_k} \right> =
\beta_1^2 \; (1 + {\cal F} \, \delta_{j k}) .
\label{w+w+}
\end{equation}

In the next step, we will average over the realizations of disorder
the corresponding expressions for MFPT and MRT, up to first order 
in the bias parameter $\epsilon$. For this model, in both problems 
we need to use the following expansions
\numparts
\begin{eqnarray}
\fl
\prod_{j = \alpha}^{\beta} \frac{w^-_j}{w^+_j} \simeq
1 - a \epsilon \,\sum_{j = \alpha}^{\beta} \frac{1}{w^+_j} ,
\label{exp_a}
\\
\fl
\sum_{k = \alpha}^{x} \,\prod_{j = \alpha}^{\beta}
\frac{w^-_j}{w^+_j} \simeq
x -\alpha +1 - a \epsilon
\sum_{k = \alpha}^{x} \,\sum_{j = \alpha}^{\beta} \frac{1}{w^+_j} ,
\label{exp_b}
\\
\fl
\sum_{k = -M}^{x-1} \frac{1}{w^+_k} \,
\prod_{j = k+1}^{x} \frac{w^-_j}{w^+_j} \simeq
\sum_{k = -M}^{x-1} \frac{1}{w^+_k} - a \epsilon \,
\sum_{k = -M}^{x-1} \,\sum_{j = k+1}^{x} \frac{1}{w^+_k w^+_j} ,
\label{exp_c}
\\
\fl
\sum_{k = -M}^{x-1} \frac{1}{w^+_k} \,\sum_{k = k+1}^{x} \,
\prod_{j = k+1}^{i} \frac{w^-_j}{w^+_j} \simeq
\sum_{k = -M}^{x-1} \frac{x-k}{w^+_k} - a \epsilon \,
\sum_{k = -M}^{x-1} \,\sum_{i = k+1}^{x} \,\sum_{j = k+1}^{i}
\frac{1}{w^+_k w^+_j} .
\label{exp_d}
\end{eqnarray}
\endnumparts
We must note that in the last term of equations~(\ref{exp_c})
and ~(\ref{exp_d}) $j \neq k$. Therefore, the average of these
expressions do not involve the fluctuation of disorder.
However, when we average the equation~(\ref{MFPT:aa}),
expanding up to first order in $\epsilon$ and using
equations~(\ref{exp_a})--(\ref{exp_d}), we obtain
\begin{equation}
\fl
\left< T_n \right> \simeq \Theta_0(n,L,M) \,\beta_1
+ \left[ \Theta_1(n,L,M) + \Theta_2(n,L,M) \,{\cal F} \right]
\,a \,\beta_1^2 \,\epsilon \,,
\label{T(e):aa}
\end{equation}
where the functions $\Theta_0(n,L,M)$, $\Theta_1(n,L,M)$
and $\Theta_2(n,L,M)$ are defined in the appendix.
Here, the fluctuation of disorder ${\cal F}$ is introduced
by the expansion of the denominator of expression~(\ref{MFPT:aa}).
Following the calculations given in the appendix, we obtain
\numparts
\begin{eqnarray}
\fl
\Theta_0(n,L,M) = \frac{(L+1-n) \,(M+1+n)}{2} ,
\label{T0}
\\
\ms
\fl
\frac{\Theta_1(n,L,M)}{\Theta_0(n,L,M)} =
\frac{L-M+3-2n}{6} ,
\label{T1}
\\
\bs
\fl
\frac{\Theta_2(n,L,M)}{\Theta_0(n,L,M)} =
\frac{2L+M+3-n}{3 \,(L+M+2)} .
\label{T2}
\end{eqnarray}
\endnumparts
Therefore, the averaged MFPT results in
\begin{eqnarray}
\fl
\left< T_n \right> \simeq
\frac{(L+1-n)(M+1+n)}{2 \beta_1^{-1}}
\nonumber \\
\lo \times
\left[ 1 +
\left( \frac{L-M+3-2n}{6}+\frac{2L+M+3-n}{3(L+M+2)} 
\,{\cal F} \right)
a \,\beta_1 \,\epsilon
\right] \,.
\label{<T>:aa}
\end{eqnarray}
Taking the limit $\epsilon \rightarrow 0$ in equation~(\ref{<T>:aa}),
we arrive to the known expression for the MFPT of a homogeneous
symmetrical chain with the effective coefficient $\beta_1^{-1}$.
Note that, for this class of disorder, the asymmetry in the
hopping transitions links the strength of the bias with the
fluctuation of disorder.

On the other hand, for the residence problem we need to evaluate
the expression
\begin{equation}
\fl
\left< \sum_{k = n}^{\infty} \,\prod_{j = L+1}^{k}
\frac{w^-_j}{w^+_j} \right> \simeq \sum_{i = 0}^{\infty}
\exp \left[ - a \,\epsilon \,\beta_1 (i+n-L) \right] \simeq
\frac{1 - a \epsilon \beta_1 (n-L)}{a \epsilon \beta_1} ,
\label{series}
\end{equation}
which is valid for $n \geq L+1$.
Then, taking $0 < \epsilon \leq 1$, so that the bias field points
to the right, and using equations~(\ref{exp_c}),~(\ref{exp_d})
and~(\ref{series}), from equation~(\ref{tau>L})--(\ref{tauinD}) we
obtain the averaged MRT
\begin{equation}
\fl
\left <\tau_n \right> \simeq
\frac{L+M+1}{a \,\epsilon} \left\{
\begin{array}{ll}
1
& \mbox{for $n < -M$} \\
\displaystyle
1 - \frac{(n+M) (n+M+1)}{2 \,(L+M+1)} \,a \,\beta_1 \,\epsilon
& \mbox{for $-M \leq n \leq L$} \\
\displaystyle
1 - \left[ n - (L - M) / 2 \right] \,a \,\beta_1 \,\epsilon
& \mbox{for $n >  L$}  \;.
\end{array}
\right.
\label{<tau>}
\end{equation}
This expression is equivalent to the diffusive regime of 
a homogeneous chain. Therefore, it can be obtained from
equation~(\ref{tauord}) taking $\gamma = 1 - \epsilon$ 
and expanding up to first order in $\epsilon$.
The effect of disorder appears as a renormalization of the parameters
$a$ and $\epsilon$ according to $a \rightarrow \beta_1^{-1}$ and 
$\epsilon \rightarrow a \,\beta_1 \,\epsilon$~\cite{Pury02}.  
Note that, in the limit $\epsilon \rightarrow 0$, from
equation~(\ref{<tau>}) we obtain a divergence. This is a direct
consequence of the fact that the MRT is not a defined quantity 
for unbiased chains.

We remark that the results given by equations~(\ref{<T>:aa})
and~(\ref{<tau>}) are valid for weak disorder.
When the quantities $\beta_k$ are not finite,
we have to deal with a perturbative approach
for analyse the behavior of the averaged MFPT
and MRT~\cite{Pury02}.

%%%%%%%%%%%%%%%%%%%%%%%%%%%%%%%%%%%%%%%%%%%%%%%%%%%%%%%%%%%%%%%%%%%%%
\section{Multiplicative model}
\label{sec:mult}
%%%%%%%%%%%%%%%%%%%%%%%%%%%%%%%%%%%%%%%%%%%%%%%%%%%%%%%%%%%%%%%%%%%%%

Our second asymmetric random-site model is defined by
$w_j^+ = a \,\eta_j$ and $w_j^- = b \,\eta_j$,
where $a$ and $b$ are positive constants and $\{ \eta_j \}$
are taken independent but identically distributed positive
random variables with $\langle \eta_j \rangle = 1$.
Thus, we obtain $w_j^- / w_j^+ = b/a = \gamma$.
It is immediate to recognize that the averaged expressions
for MFPT and MRT are obtained replacing $1/w^+_k$ by
$\beta_1 \equiv \left< 1 / \left( a \,\eta_j \right) \right>$
in all the terms of equations~(\ref{MFPT:aa})
and~(\ref{tau>L})--(\ref{tauinD}).
Therefore, the averaged formulae for this model of disorder
corresponds to the homogeneous case, given in
section~\ref{sec:order}, with the substitution 
$a \rightarrow \beta_1^{-1}$.
We must stress that the averaged expressions obtained by this way
are exact expressions for all value of the bias parameter $\gamma$.
Particularly, taking $\gamma = 1 - \epsilon$, up to first order in
$\epsilon$, we obtain for the averaged MFPT
\begin{equation}
%\fl
\left< T_n \right> \simeq
\frac{(L+1-n)(M+1+n)}{2 \beta_1^{-1}}
\times
\left[ 1 +
\frac{L-M+3-2n}{6} \,\epsilon
\right] \,.
\label{<T>mult}
\end{equation}
As a notable remark, we observe that there is not coupling
between the bias $\epsilon$ and the fluctuation of disorder
${\cal F}$ for the multiplicative model.
We emphasize that the multiplicative asymmetric disordered
model cannot be easily worked out by using FEMA~\cite{HC90,Pury94}
or in general using a perturbative method~\cite{Pury02}.

%%%%%%%%%%%%%%%%%%%%%%%%%%%%%%%%%%%%%%%%%%%%%%%%%%%%%%%%%%%%%%%%%%%%%
\section{One reflecting boundary}
\label{sec:cut}
%%%%%%%%%%%%%%%%%%%%%%%%%%%%%%%%%%%%%%%%%%%%%%%%%%%%%%%%%%%%%%%%%%%%%

In this section, we consider a reflecting boundary condition 
at the left extreme of the interval. The reflecting boundary 
can be modeled by setting $w_{-M}^- = w_{-(M+1)}^+ = 0$.
Without lost of generality, we take $M=0$, therefore our interval
of interest is $[0,L]$.
From equation~(\ref{MFPT:aa}), taking $M=0$ and $w_{0}^- = 0$,
we immediatly obtain for the MFPT
\numparts
\begin{eqnarray}
T_0 &=&
\displaystyle\sum_{k=0}^{L} \,\frac{1}{w^+_k} +
\displaystyle\sum_{k=0}^{L-1} \,\frac{1}{w^+_k} \,
\sum_{i=k+1}^{L} \prod_{j=k+1}^{i} \;\frac{w^-_j}{w^+_j} \;,
\label{MFPT_0:ra}
\\
T_1 &=& T_0 - \frac{1}{w^+_0} \;,
\label{MFPT_1:ra}
\\
T_n &=& T_0 -
\sum_{k=0}^{n-1} \,\frac{1}{w^+_k} -
\sum_{k=0}^{n-2} \,\frac{1}{w^+_k} \,
\sum_{i=k+1}^{n-1} \prod_{j=k+1}^{i} \;\frac{w^-_j}{w^+_j}
\;\; \mbox{for $2 \leq n \leq L$} \,.
\label{MFPT:ra}
\end{eqnarray}
\endnumparts
Equation~(\ref{MFPT_0:ra}) was reported for discrete time random
walks by Murthy and Kehr~\cite{MK89}.
The main effect of the reflecting boundary is the disappearance of
the denominator in the expression of the MFPT.
For the homogeneous (nondisordered) case, for which
$w^+_j = a$, $w^-_j = b$ and $\gamma = b/a$, we obtain for
$0 \leq n \leq L$ that
\begin{equation}
T_n = \displaystyle\frac{1}{a (1-\gamma)}
      \left( L+1-n -
      \frac{\gamma^{n+1}-\gamma^{L+2}}{1 - \gamma}
      \right) \,.
\label{Tord:ra}
\end{equation}
The small bias expansion ($\gamma = 1 - \epsilon$) of
equation~(\ref{Tord:ra}) is given by
\begin{equation}
T_n \simeq
\displaystyle\frac{(L+1)(L+2) - n(n+1)}{2 \,a}
\left[ 1 -
\left( \frac{L}{3} + \frac{n(n+1)}{3\,(L+2+n)} \right)
\epsilon
\right] \,.
\label{Tweak:ra}
\end{equation}

In the guideline of section~\ref{sec:weak}, we can compute
the averaged MFPT for a walker in a chain with reflecting-absorbing
boundary conditions and additive disorder.
In this manner, we obtain up to first order in $\epsilon$
\begin{equation}
\fl
\left< T_n \right>  \simeq
\frac{(L+1)(L+2) - n(n+1)}{2 \,\beta_1^{-1}}
\left[ 1 -
\left( \frac{L}{3} + \frac{n(n+1)}{3\,(L+2+n)} \right)
a \,\beta_1 \,\epsilon
\right] \,.
\label{<T>:ra}
\end{equation}
Equation~(\ref{<T>:ra}) is exactly the same as~(\ref{Tweak:ra})
for weak biased nondisordered chains if we set
$a \rightarrow \beta_1^{-1}$ and
$\epsilon \rightarrow a \,\beta_1 \,\epsilon$.
Strikingly, for reflecting-absorbing boundaries the fluctuation
of disorder is not present in the averaged MFPT.
From equation~(\ref{Tord:ra}) and~(\ref{Tweak:ra}), the same argument
presented in section~\ref{sec:mult} allows us to reckon the averaged
MFPT for a disordered chain under multiplicative disorder,
substituting $a$ by $\beta_1^{-1}$.

The presence of a reflecting boundary at the left of the interval
has not effect on the RT problem when the bias points to the right.
This fact can be easily seen from
equations~(\ref{tau>L})--(\ref{tauinD}), where we found that MRT's
expressions do not depend on the hopping transitions $w_{-M}^-$
and $w_{-(M+1)}^+$.
However, if the reflecting boundary is at the right of the interval,
when the bias points to the right, the MRT diverges as is expected.
This fact can be seen from equations~(\ref{tau_L})--(\ref{tauinD})
taking $w_{L+1}^- = 0$ and the limit $w_{L}^+ \rightarrow 0$ 
(see figure~\ref{RT-Fig}).

%%%%%%%%%%%%%%%%%%%%%%%%%%%%%%%%%%%%%%%%%%%%%%%%%%%%%%%%%%%%%%%%%%%%%
\section{Concluding remarks}
\label{sec:fin}
%%%%%%%%%%%%%%%%%%%%%%%%%%%%%%%%%%%%%%%%%%%%%%%%%%%%%%%%%%%%%%%%%%%%%

We have presented an algebraic method for calculating MFPT and MRT
of a random walk on one-dimensional lattices for quenched disorder.
The starting points are the one step equations~(\ref{eq_T})
and~(\ref{eq_tau}).
We have obtained the exact solution~(\ref{MFPT:aa}) for
the MFPT and the exact expressions~(\ref{tau<-M}) and
(\ref{tau>L})--(\ref{tauinD}) for the MRT.
Also, we have considered a reflecting boundary
in the section~\ref{sec:cut}.
Two models for site disorder in the chain were considered,
namely, additive and multiplicative.
The expressions for MFPT and MRT were exactly averaged for 
both kinds of disorder.
The main difference between these models is in the coupling of
the inverse moments of jump transitions and the bias parameter
in the averaged quantities.
For the additive model of section~\ref{sec:weak}, the terms
to first order in the bias parameter are proportional to
$a \,\beta_1 \,\epsilon$ 
[see equations~(\ref{<T>:aa}),~(\ref{<tau>}) and~(\ref{<T>:ra})].
Moreover, for absorbing-absorbing boundaries, the bias links the
strength parameter $\epsilon$ with the fluctuation of disorder 
${\cal F}$ in the averaged MFPT [see equation~(\ref{<T>:aa})].  
On the other hand, for the multiplicative model, the terms
to first order in the bias parameter are proportional to
$a \,\epsilon$ and there is not any kind of linking between 
$\epsilon$ and ${\cal F}$ in the averaged MFPT.

In this work we have considered models of site disorder.
The pair of transition 
probabilities $w^+_j$ and $w^-_j$,
associate to the site $j$, has the same distribution
through the random variable $\xi_j$ or $\eta_j$.
In addition, the exact equation (\ref{MFPT_bond})
allow us to consider models of bond disorder.
In this case, the pair of jump probabilities
$w^+_j$ and $w^-_{j+1}$, associate to the bond
between the sites $j$ and $j+1$, has the same
distribution.
For tackle this problem, we only need to rewrite
the sums in the prefactor in the following manner:
\begin{equation}
\sum_{k=-M}^{x} \prod_{j=-M}^{k} \frac{w^-_j}{w^+_j} =
w^-_{-M} \left( \frac{1}{w^+_{-M}} +
\sum_{k=-M}^{x-1} \frac{1}{w^+_{k+1}} \prod_{j=-M}^{k}
\frac{w^-_{j+1}}{w^+_j} \right)
\label{Sum_bond}
\end{equation}
Similar arranging can be made with the residence time expressions
to deal with bond disordered problems.

%%%%%%%%%%%%%%%%%%%%%%%%%%%%%%%%%%%%%%%%%%%%%%%%%%%%%%%%%%%%%%%%%%%%%
%%% Acknowledgments
%%%%%%%%%%%%%%%%%%%%%%%%%%%%%%%%%%%%%%%%%%%%%%%%%%%%%%%%%%%%%%%%%%%%%
%
\ack
This work was partially supported by the
``Se\-cre\-ta\-r\'\i a de Cien\-cia y Tec\-no\-lo\-g\'\i a
de la Uni\-ver\-si\-dad Na\-cio\-nal de C\'or\-doba''
under Grant No. 05/B160 (Res.\ SeCyT 194/00).

%%%%%%%%%%%%%%%%%%%%%%%%%%%%%%%%%%%%%%%%%%%%%%%%%%%%%%%%%%%%%%%%%%%%%
\appendix
\section*{Appendix}
\label{sec:app}
\setcounter{section}{1}
%%%%%%%%%%%%%%%%%%%%%%%%%%%%%%%%%%%%%%%%%%%%%%%%%%%%%%%%%%%%%%%%%%%%%

In order to perform the average over disorder for the additive 
model of section~\ref{sec:weak}, we need previously to evaluate 
the following functions:
\begin{eqnarray}
\fl
f_0(x,M) \equiv \sum_{k=-M}^{x-1} (x-k)
%= x \,(x+M) - \frac{1}{2} \left[ x \,(x-1) - M \,(M+1) \right] ,
= \frac{(x+M)\,(x+M+1)}{2} \,,
\label{f0}
\\
\bs
\fl
f_1(x,M) \equiv
\sum_{k=-M}^{x-1} \,\sum_{i=k+1}^{x} \,\sum_{j=k+1}^{i} \,1
%\nonumber\\
%\ms
%\lo= \frac{1}{2}
%\left\{
%(x+M) \,(x^2+x)
%- \frac{(2x+1)}{2} \left[ x \,(x-1) - M \,(M+1) \right]
%\right.
%\nonumber\\
%\ms
%\left.
%+ \frac{1}{6} \left[ x \,(x-1) \,(2x-1) + M \,(M+1) \,(2M+1) \right]
%\right\} ,
= \frac{(x+M)\,(x+M+1)\,(x+M+2)}{6} \,,
\label{f1}
\\
\bs
\fl
f_2(x,M) \equiv
\sum_{k=-M}^{x-1} \,\sum_{i=k+1}^{x} \,\sum_{j=k+1}^{i}
\delta_{jk} = 0 ,
\label{f2}
\\
\bs
\fl
f_3(x,L,M) \equiv
\sum_{k=-M}^{L} \,\sum_{i=-M}^{x} \,\sum_{j=-M}^{i} \,1
%\nonumber\\
%\ms
%\lo= \frac{(L+M+1)}{2} \,
%\left[ x \,(x+1) - M \,(M+1) + 2 \,(x+M+1) (M+1) \right] ,
= \frac{(L+M+1)\,(x+M+1)\,(x+M+2)}{2} \,,
\label{f3}
\\
\bs
\fl
f_4(x,L,M) \equiv
\sum_{k=-M}^{L-1} \,\sum_{i=-M}^{x} \,\sum_{j=-M}^{i} \,(L-k)
\nonumber\\
\ms
\lo=
\frac{(L+M)\,(L+M+1)\,(x+M+1)\,(x+M+2)}{4} \,,
\label{f4}
\\
\bs
\fl
f_5(x,L,M) \equiv
\sum_{k=-M}^{L} \,\sum_{i=-M}^{x} \,\sum_{j=-M}^{i} \,\delta_{jk}
= \frac{(x+M+1)\,(x+M+2)}{2} \,,
\label{f5}
\\
\bs
\fl
f_6(x,L,M) \equiv
\sum_{k=-M}^{L-1} \,\sum_{i=-M}^{x} \,\sum_{j=-M}^{i}
\,(L-k) \,\delta_{jk}
\nonumber\\
\ms
\lo=
%(x+M+1) \,\left[ L \,(M+1) + \frac{1}{2}\,M \,(M+1) \right]
%\nonumber\\
%+ \frac{1}{2} \,\left( L - \frac{1}{2} \right) \,
%\left[ x \,(x+1) - M \,(M+1) \right]
%\nonumber\\
%\ms
%-\frac{1}{12} \left[ x \,(x+1) \,(2x+1) + M \,(M+1) \,(2M+1) \right]
\frac{(3L+2M-x)\,(x+M+1)\,(x+M+2)}{6} \,.
\label{f6}
\end{eqnarray}
Using the functions defined above, the functions
$\Theta_0(n,L,M)$, $\Theta_1(n,L,M)$ and $\Theta_2(n,L,M)$, 
that appear in equation~(\ref{T(e):aa}), are defined by
\begin{eqnarray}
\fl
\Theta_0(n,L,M) \equiv \frac{n+M+1}{L+M+2}
\left[ (L+M+1) + f_0(L,M) \right]
\nonumber\\
\ms
-
\left[ (n+M) + f_0(n-1,M) \right] ,
\label{theta0}
\\
\bs
\fl
\Theta_1(n,L,M) \equiv \frac{n+M+1}{L+M+2}
\left[
\frac{1}{L+M+2} \,\left( f_3(L,L,M) + f_4(L,L,M) \right)
\right.
\nonumber\\
\ms
\left.
- \frac{1}{n+M+1} \left( f_3(n-1,L,M) + f_4(n-1,L,M) \right)
- f_1(L,M)
\right]
\nonumber\\
\ms
+ f_1(n-1,M) ,
\label{theta1}
\\
\bs
\fl
\Theta_2(n,L,M) \equiv \frac{n+M+1}{L+M+2}
\left[
\frac{1}{L+M+2} \,\left( f_5(L,L,M) + f_6(L,L,M) \right)
\right.
\nonumber\\
\ms
\left.
- \frac{1}{n+M+1} \left( f_5(n-1,L,M) + f_6(n-1,L,M) \right)
\right] .
\label{theta2}
\end{eqnarray}
Equations~(\ref{T0})--(\ref{T2}) follow from
equation~(\ref{theta0})--(\ref{theta2}), respectively,
replacing the functions $f_i$ by their explicit expressions.

\newpage
%%%%%%%%%%%%%%%%%%%%%%%%%%%%%%%%%%%%%%%%%%%%%%%%%%%%%%%%%%%%%%%%%%%%%
\section*{References}
%%%%%%%%%%%%%%%%%%%%%%%%%%%%%%%%%%%%%%%%%%%%%%%%%%%%%%%%%%%%%%%%%%%%%
%

%%%%%%%%%%%%%%%%%%%%%%%%%%%%%%%%%%%%%%%%%%%%%%%%%%%%%%%%%%%%%%%%%%%%%
\end{document}